\documentclass{aa}

\usepackage{graphicx}
\usepackage{txfonts}
\begin{document}

   \title{Light Walls Around Sunspots Observed by the Interface Region Imaging Spectrograph}

   \author{Y. J. Hou \inst{1,2}, T. Li \inst{1,2}, S. H. Yang \inst{1,2}
          \and
          J. Zhang\inst{1,2}
          }

   \institute{Key Laboratory of Solar Activity, National Astronomical Observatories,
   Chinese Academy of Sciences, Beijing 100012, China\\
             \email{yijunhou@nao.cas.cn}
             \and
             University of Chinese Academy of Sciences, Beijing 100049, China
             }

   \date{Received ????; accepted ????}


  \abstract
   {The Interface Region Imaging Spectrograph (\emph{IRIS}) mission provides high-resolution
   observations of \textbf{the} chromosphere and transition region. Using these data, some
   authors have reported the new finding of light walls above sunspot light bridges.
   }
   {We try to determine whether the light walls exist somewhere else \textbf{in active regions besides light bridges}
   and examine the evolution of these walls' materials.
   }
   {Employing half-year (from 2014 December to 2015 June) high tempo-spatial data from the \emph{IRIS},
   we find lots of light walls either around sunspots or above light bridges.
   }
   {For the first time, we report one light wall near an \textbf{umbral-penumbral} boundary and
   another along a neutral line between two small sunspots. The former light wall \textbf{posses a}
   multilayer structure and is associated with the emergence of positive magnetic flux
   among the ambient negative field. The latter light wall is relevant to a filament activation
   and the wall body consists of the filament material, which flowed to a \textbf{remote plage region
   with a negative magnetic field} after the light wall disappeared.
   }
   {\textbf{These new observations reveal that these light walls are multi-layer and multi-thermal
   structures which occur along magnetic neutral lines in active regions.}
   }


   \titlerunning{Light wall}
   \authorrunning{Hou et al.}

   \maketitle
%

\section{Introduction}

Sunspots are concentrations of magnetic fields and appear as dark patches on the visible
solar surface. Their strong magnetic fields which are up to 4000 G inhibit \textbf{the energy's
normal convective transports in the convection zone} (Solanki 2003; Thomas \& Weiss 2004).
\textbf{In the layers of photosphere and chromosphere, a typical sunspot is characterized by a dark
core, the umbra, and a less dark halo, the penumbra, surrounding the umbra. Based on the work (Curdt et al. 2001)
about the sunspot spectra obtained by SUMER (Solar Ultraviolet Measurements of Emitted Radiation), Tian et al. (2009) revealed that the
transition region above sunspots are higher and more extended in comparison to the plage regions.}
In the sunspot umbra, the overturning motion of the plasma is
hindered by the strong magnetic field, thus leading to a lower temperature in the photospheric
layers due to the reduced energy input from below (Gough \& Tayler 1966).
Bright structures within the umbra are signatures of not completely suppressed convection
and light bridges are the best known representative of these structures (Sobotka et al. 1993).
The magnetic field of light bridges is generally weaker and more inclined
than the local strong and vertical field, forming a magnetic canopy (Lites et al. 1991;
Rueedi et al. 1995; Leka 1997; Jur{\v c}{\'a}k et al. 2006).

Recently, Yang et al. (2015) reported an oscillating light wall above a sunspot light bridge
with the high tempo-spatial data from the \emph{Interface Region Imaging Spectrograph}
(\emph{IRIS}; De Pontieu et al. 2014). In their work, the light wall is brighter than the
ambient areas while the wall top is much brighter than the wall body in 1330 {\AA} passband.
Similar observations of plasma ejections from a light bridge have also been reported previously
(Asai et al. 2001; Shimizu et al. 2009; Robustini et al. 2015). Nevertheless, many questions
about the light wall are still \textbf{waiting} to be answered, for example, can a light wall only be rooted on
a light bridge? And \textbf{what physical mechanism drives the kinematic evolution of a light wall?}

Our work mainly concerns the structures of light walls at different locations and the kinematics
of the material in the walls. Checking about half year's \emph{IRIS} data from 2014 December to 2015 June, we notice
that most of the light walls are rooted above the light bridges, but some light walls are observed
near the \textbf{umbral-penumbral} boundary, and a light wall is located along the polarity inversion line.
Using \textbf{coordinated observations} from the \emph{IRIS}, the \emph{Solar Dynamic Observatory}
(\emph{SDO}; Pesnell et al. 2012) and the New Vacuum Solar Telescope (NVST; Liu et al. 2014)
of the \emph{Fuxian Solar Observatory} in China, we report \textbf{on} two light walls around sunspots in detail.

\section{Observations and Data Analysis}

In this Letter, one series of \emph{IRIS} slit-jaw 1330 {\AA} images (SJIs) and two series
of SJIs in 1400 {\AA} are adopted, \textbf{and they are all Level 2 data}. The 1330 {\AA} passband contains
emission from the strong C II 1334/1335 {\AA} lines formed in the upper chromosphere and transition region
while the 1400 {\AA} channel contains emission from the Si IV 1394/1403 {\AA} lines formed in lower
transition region.
On 2014 December 19, the SJIs in 1330 {\AA} focused on
NOAA AR 12242 were taken from 17:32:53 UT to 18:30:33 UT with a pixel scale of 0.{\arcsec}333 and
a cadence of 12 s. For this event's spectroscopic analysis, we use the line of Si IV in 1403 {\AA}
which is formed in the middle transition region with a temperature of about 10$^{4.9}$ K (Li et al. 2014;
Tian et al. 2014). \textbf{The spectral data are taken in a large coarse 8-step mode with 12 s cadence.}
Since the Si IV profiles are close to Gaussian distribution, we use
single-Gaussian \textbf{fits} to approximate the 1403 {\AA} line (Peter et al. 2014).
On 2015 January 16, the 1400 {\AA} SJIs focused on NOAA 12259 were obtained
from 07:04:41 UT to 08:04:20 UT and from 08:42:12 UT to 09:21:06 UT with a cadence of 13 s and
a pixel size of 0.{\arcsec}333.

The NVST pointed at NOAA 12259 on 2015 January 16 as well and we got a
series of H$\alpha$ images. From 07:18:00 UT to 08:14:00 UT, the H$\alpha$ 6562.8 {\AA}
observations had a cadence of 12 s and a pixel scale of 0.{\arcsec}164. The Level 0 H$\alpha$
images are firstly calibrated to Level 1 with dark current subtracted and flat field corrected,
and then the Level 1 images are reconstructed to Level 1+ by speckle masking (see Weigelt 1977;
Lohmann et al. 1983).

Moreover, the Atmospheric Imaging Assembly (AIA; Lemen et al. 2012) and the Helioseismic and
Magnetic Imager (HMI; Scherrer et al. 2012) observations from the \emph{SDO} have been used as well.
The 304 {\AA}, 171 {\AA} and 193 {\AA} data with a cadence of 12 s and a pixel size of 0.{\arcsec}6 are
exhibited to show the light wall in different temperatures. The observations of 1600 {\AA} from
2015 January 15 15:00:00 UT to January 16 15:00:00 UT have been taken to replace the 1400 {\AA} SJIs
for a long-time measurement of brightness around light wall base. We also use the full-disk line-of-sight
(LOS) magnetogram and the intensitygram from the HMI, with a cadence of 45 s and a sampling of
0.{\arcsec}5 pixel$^{-1}$. Using the cross-correlation method, we co-align all the \emph{IRIS},
NVST, and \emph{SDO} images according to specific features (Yang et al. 2014).

\section{Results}
\subsection{A Light Wall with Multilayer Structure Along an \textbf{Umbral-Penumbral} Boundary}
\begin{figure}
\centering
\includegraphics [width=0.46\textwidth]{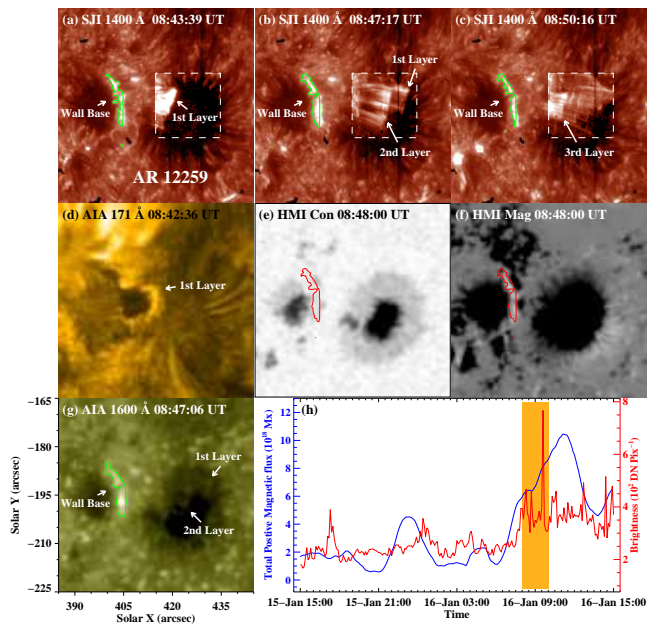}
\caption{
Panels (a)-(c): \emph{IRIS} 1400 {\AA} images showing the development of the light wall on 2015 January 16.
The green curves in different panels are the contours of the brightness at different times, which denote the bases of light walls approximately.
\textbf{The features in white dashed windows are sharpened.}
Panels (d) and (g): AIA 171 {\AA} and 1600 {\AA} images exhibiting the light wall in two different temperatures.
Panels (e)-(f): HMI continuum intensity and LOS magnetogram displaying the light wall's location around sunspots and
the underlying magnetic field, respectively. The red curves are the duplications of the contour in panel (b).
Panel (h): positive magnetic flux (blue curve) and brightness (red curve) of the wall base's location during 24 hours.
The orange region marks the stage of light wall's appearance.
\textbf{The full temporal evolution of the 1400 {\AA}, 171 {\AA}, 304 {\AA} and 1600 {\AA} images is available
as a movie (1.mp4) in the online edition.}
}
   \label{fig1}
\end{figure}

A light wall along an \textbf{umbral-penumbral} boundary appeared on 2015 January 16 around a sunspot of
NOAA 12259. Figures 1(a)-1(c) display \textbf{the light wall possessed a
multilayer structure (see movie attached to Figure 1)}. In the 1400 {\AA} images, we can see that the wall base
(green contours in panels (a)-(c)) is brighter than the surrounding area. At about 08:43 UT, \textbf{the first and the second
layers of the light wall emerged from the \textbf{umbral-penumbral} boundary but they separated and extended different distances
several minutes later}. Figure 1(b) shows the clear scene of two light wall layers' existence
at 08:47:17 UT. When the two layers fell back to their wall base, the third layer appeared (panel (c)).
Panel (d) exhibits the first layer of the wall in 171 {\AA} and it is clear that the layer top has
high emission while the emissions at both the base and body are quite low. The HMI intensitygram in
panel (e) shows the location of wall base along the \textbf{umbral-penumbral} boundary. To understand the magnetic
field environment around the wall base, we measure the positive magnetic flux (see the blue line in panel (h))
and the brightness (red line) at the wall base's location.
Since the data from \emph{IRIS}/SJI 1400 {\AA} couldn't cover the whole period for brightness measurement,
we take \emph{SDO}/AIA 1600 {\AA} observations \textbf{from 15-Jan 15:00 UT to 16-Jan 15:00 UT} for substitution (panel (g)).
It indicates that a \textbf{line-like} positive magnetic field emerged underneath the wall base (panel (f)). The emerging flux
underwent two peaks during 24 hours, and the light wall appeared at the rising phase of the second peak
(orange region in panel (h)). Meanwhile, the brightness at the wall base increased evidently.

\begin{figure}
\centering
\includegraphics [width=0.46\textwidth]{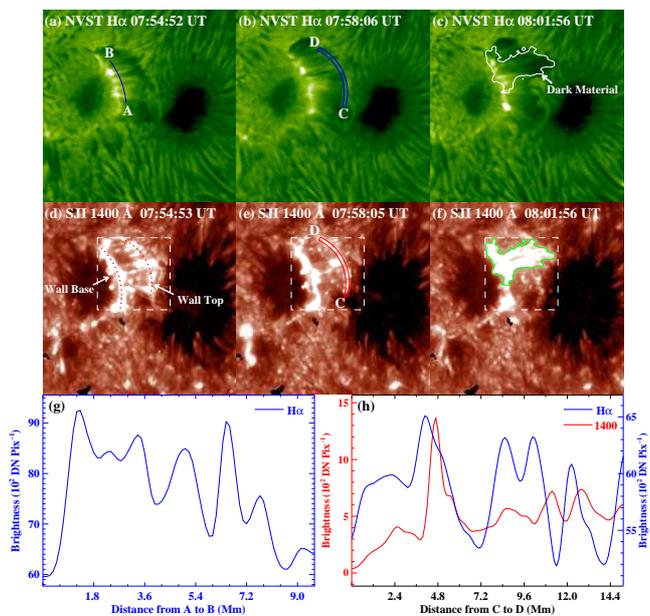}
\caption{
Panels (a)-(c): NVST H$\alpha$ images displaying the light wall one hour earlier than that in Figure 1.
Panels (d)-(f): a series of \emph{IRIS} 1400 {\AA} images showing the light wall. The green curve in panel (f) is the
contour of a brightening patch, and the white curve in panel (c) is the duplication of the patch contour here.
\textbf{The light walls in white dashed windows are sharpened.}
Panel (g): H$\alpha$ brightness along curve ``A-B'' in panel (a).
Panel (f): H$\alpha$ (blue curve) and 1400 {\AA} (red curve) brightness within arc-sector domain ``C-D'' in panels (b) and (e) respectively.
\textbf{An animation (2.mp4) of the 1400 {\AA} and the H$\alpha$ images is available online.}
}
\label{fig2}
\end{figure}

To study the multi-wavelength appearances of the light wall, we employ the NVST H$\alpha$ data and compare them
with corresponding \emph{IRIS} SJIs of 1400 {\AA} (see Figure 2). The H$\alpha$
observations only cover the period from 07:18:00 UT to 08:14:00 UT \textbf{(see movie attached to Figure 2)}, which is one hour earlier than
that shown in Figure 1, but fortunately, the simultaneous \emph{IRIS} data are also available and the
light wall at this time is still conspicuous. At 07:54:52 UT, the light wall in H$\alpha$ seemed to
be composed of multiple bright and dark threads. However, the 1400 {\AA} image only displays the bright
threads while the dark threads could not be observed. To research the fine structures of the light wall,
we make a cut along the slice ``A-B'' (see the blue curve in panel (a)) and present the brightness along the cut
in panel (g). To compare the emission between H$\alpha$ and 1400 {\AA} images, arc-sector domains (see the blue and
red curves in panels (b) and (e)) along the light wall's top are selected and the brightness along the arc-sector domains
are shown in panel (h). There is almost no correlation between the blue curve and the red one, except the first peak around 4.8 Mm.
At 08:01:56 UT, a brightening patch of the light wall was observed in 1400 {\AA} and we
make a contour (green curve in panel (f)) to outline it. But in the H$\alpha$ image of panel (c), the brightening
could not be observed at this region, instead, a bulk of dark materials occupied the area.

\subsection{A Light Wall Associated with Filament Activation Above a Neutral Line}
\begin{figure}
\centering
\includegraphics [width=0.46\textwidth]{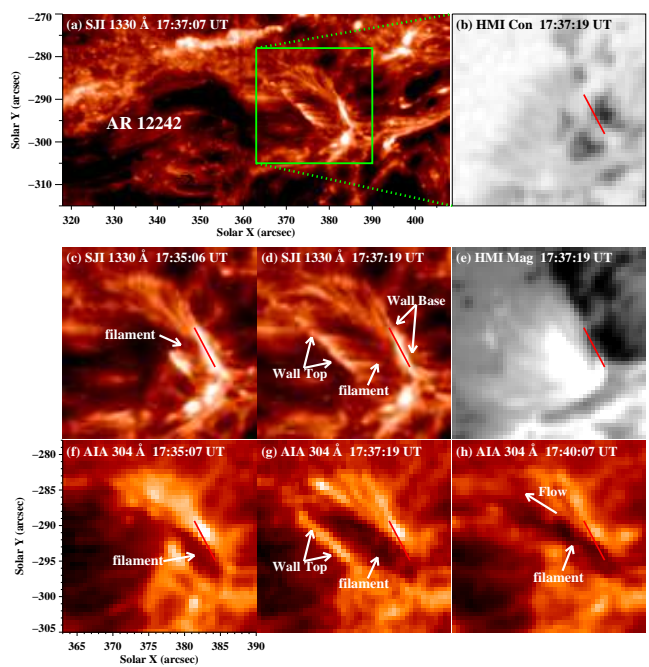}
\caption{
Panel (a): \emph{IRIS} SJI of 1330 {\AA} showing the light wall on 2014 December 19.
The green square outlines the field of view of panels (b)-(h).
Panel (b): \emph{SDO}/HMI continuum intensity displaying the light wall's location around sunspots. The red line means the base
of light wall which is rooted on the neutral line between two small sunspots.
\textbf{Panels (c)-(d) and (f)-(h): sequences of 1330 {\AA} and 304 {\AA} images displaying the evolution of the light wall
and the associated filament.}
Panel (e): HMI LOS magnetogram revealing the magnetic field beneath the light wall.
\textbf{The temporal evolution of the 1330 {\AA}, 193 {\AA} and 304 {\AA} images is available as a movie (3.mp4) online.}
}
\label{fig3}
\end{figure}

The light wall above a neutral line occurred on 2014 December 19 and was associated with a filament activation
\textbf{(see movie attached to Figure 3)}. Figure 3(a) shows the light wall in NOAA 12242 and it seems like a comb morphology.
According to the intensitygram from \emph{SDO}/HMI in panel (b), we notice that this light wall is rooted on
the neutral line (delineated by the red lines in panels (b) and (e)) between two small sunspots
with opposite-polarity magnetic fields. \textbf{Panels (c)-(d)} display the evolution of this light wall in 1330
{\AA} SJIs. Before the appearance of the light wall, a filament was observed near the neutral line
(panels (c) and (f)). Then the filament was activated and dark material moved upward, forming a
comb-shaped light wall (panels (d) and (g)). It seems that the light wall was traced out by the
activated material of the filament, and was composed of many fine structures as seen in
1330 {\AA} SJI. The wall top and base were brighter than the wall body in channels of both 304 {\AA} and
1330 {\AA}. \textbf{At 17:40 UT, the material of the light wall descended to the base and then the light
wall disappeared (panel (h))}. Moreover, associated with the filament activation, partial material
was brightened and moved to a \textbf{remote plage region which owns a negative magnetic field} after
the material of wall body fell to the base \textbf{(see movie attached to Figure 4)}.

\begin{figure}
\centering
\includegraphics [width=0.46\textwidth]{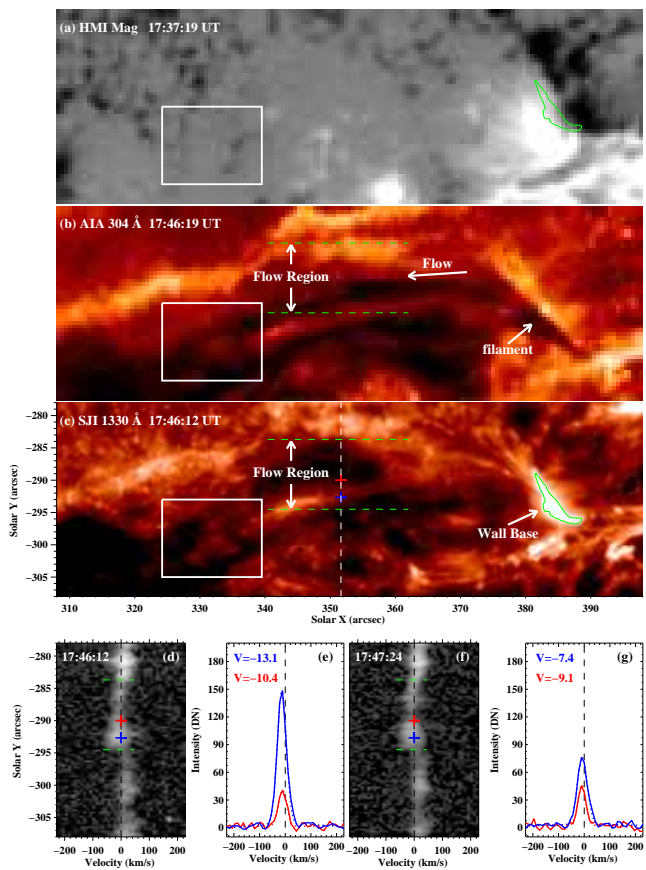}
\caption{
Panels (a)-(c): \textbf{\emph{SDO}/HMI LOS magnetogram, \emph{SDO}/AIA 304 {\AA} and \emph{IRIS} 1330 {\AA} images exhibiting the magnetic field
environment and bright flow from the base after the material of the wall body fell to the base.} The white rectangles in
panels (a)-(c) indicate the area where the flow stops. The green curve in panel (c) is brightness contour in channel 1330 {\AA} at 17:46:12 UT
and is duplicated to panel (a). The white dashed line in panel (c) shows the location of slit.
Panels (d)-(e): Si IV 1403 {\AA} spectra in the slit range of panel (c) and the profiles (solid lines) and its Gaussian fittings (dotted lines)
of this line at blue and red plus positions in panel (c).
Panels (f)-(g): similar to panels (d)-(e), \textbf{but at 17:47:24 UT}.
\textbf{An animation (4.mp4) of the 1330 {\AA} and 304 {\AA} channels shown in this figure is available in the online edition.}
}
\label{fig4}
\end{figure}

To investigate kinematic evolution of material relevant to this wall base, we choose a larger field of view
which covers the wall and the \textbf{remote plage region}. \textbf{The \emph{SDO}/HMI magnetogram, AIA 304 {\AA} and \emph{IRIS} 1330 {\AA}
images are displayed}, and thus we can study the relationship among the magnetic connectivity, light wall and bright flow from the base.
Moreover, the spectra data is employed for Doppler velocity measurement (see Figure 4). At about 17:46 UT, emission
at the wall base enhanced \textbf{in 1330 {\AA} passband} and then plenty of bright material began to move forming a flow
towards the \textbf{remote plage region} (white rectangles in panels (b) and (c)). The magnetogram in panel (a)
reveals that the magnetic fields in the rectangle are negative.
The \emph{IRIS} slit (white dashed line in panel (c)) \textbf{was located} in the middle of the material flow, so
we take the spectra data to measure the Doppler velocity of the flow and show them in panels (d)-(g). In the ``flow
region'', the profile of Si IV 1403 line is blueshifted (between two green dashed lines in panels (d) and (f)),
which means that this region is full of the flow which moves from the wall base to the rectangle area. Along the slit,
we select two locations in the flow region and show their spectra profiles and the \textbf{Gaussian fits} in panels
(e) and (g). At 17:46:12 UT, the blueshift velocities at blue and red plus positions in panel (c) are respectively 13.1
and 10.4 km s$^{-1}$ (see panel (e)). And at 17:47:24 UT, the blueshift velocities at blue and red pluses in \textbf{panel} (c)
are 7.4 and 9.1 km s$^{-1}$ respectively (see panel (g)).

\section{Conclusions and Discussion}
Employing the high tempo-spatial \emph{IRIS} observations, we find many light walls. In \emph{IRIS} 1400 {\AA} and
1330 {\AA} SJIs, the light wall is brighter than the ambient area while its base and top are much brighter than the
wall body. For the first time, we observe a light wall along the polarity inversion line between two
small sunspots in NOAA 12242. This light wall is relevant to a filament activation and the wall body
is filled with the filament material. Besides, along the \textbf{umbral-penumbral} boundary of a sunspot in NOAA 12259,
we observe a light wall \textbf{which posses a} multilayer structure and is rooted on an emerging magnetic field with
a \textbf{line-like} shape.

The light wall on 2015 January 16 appeared along the \textbf{umbral-penumbral} boundary and its base was rooted
on an emerging magnetic field. It is widely believed that the bright knots and frequent mass ejections
in the chromosphere are driven by emerging flux which reconnects with the surrounding area in succession
(Kurokawa \& Kawai 1993; Asai et al. 2001; Zhang \& Wang 2002; Bharti et al. 2007; Shimizu et al. 2009).
Different from the works of Asai et al. (2001), Shimizu et al. (2009) and Yang et al. (2015), the light
walls we report are not above light bridges and they behave as a whole rather than separate ejections.
As for the wall's multilayer structure along the \textbf{umbral-penumbral} boundary, we put forward two possible
explanations: the light wall owns just a single fan-shaped magnetic structure with continuous
perturbation or the light wall simply owns multiple fan-shaped magnetic structures.
For further study of this light wall, NVST H$\alpha$ observations are added
and we notice that the same light wall displays different features in \emph{IRIS} SJIs and H$\alpha$ images
(see Figure 2). It seems that the light wall consists of warm (1330 and 1400 {\AA}) and cold (H$\alpha$) material
simultaneously. In the channels of 1330 {\AA} and 1400 {\AA}, the cold material seems semitransparent and
optically thin  while the hot structure could be detected clearly (Li \& Zhang 2015). So the wall body
appears as brightening (see panels (d)-(f)). In H$\alpha$ passband, the cold material absorbs the radiation
strongly (Heinzel et al. 2001), then the wall body appears as dark structure (see panels (a)-(c)).
\textbf{The observations of both the NVST H$\alpha$ and \emph{IRIS} UV channel are all sensitive to chromosphere and transition region
temperatures, and the base of the light wall is brightening in these channels (see Figure 2). Moreover, the brightening is
associated with magnetic flux emergence at photosphere (see Figure 1). So we suggest that the magnetic reconnection, which may
trigger the brightening at the wall base, occurs in the lower atmosphere.}

The light wall relevant to a filament is along the polarity inversion line. And we detect a
material flow from the light wall base to a \textbf{remote plage region} which owns a negative magnetic field after the wall
body material falls to the base. By examining the magnetic field environment and the flow, we suggest that
this light wall traces part of the magnetic structure which is rooted at the positive-polarity fields side of
the neutral line. And the flow moves along the loops connecting the positive field beneath wall base and the remote negative field.

\begin{acknowledgements}
\textbf{We thank the referee for his/her valuable suggestions.}
The data are used courtesy of \emph{IRIS}, \emph{SDO} and NVST science teams. \emph{IRIS} is a NASA small
explorer mission developed and operated by LMSAL with mission operations executed at NASA Ames Research
center and major contributions to downlink communications funded by ESA and the Norwegian Space Centre.
This work is supported by the National Natural Science Foundations of China (11533008, 11303050,
11303049, 11373004, 11203037 and 11221063), the Strategic Priority Research Program$-$The Emergence
of Cosmological Structures of the Chinese Academy of Sciences (Grant No. XDB09000000) and the Youth
Innovation Promotion Association of CAS (2014043).
\end{acknowledgements}

%
%

\clearpage

\end{document}